\documentclass[sigconf]{acmart}

\usepackage{graphicx}
\usepackage{float}
\usepackage{booktabs} 
\usepackage{enumitem,kantlipsum}
\usepackage{threeparttable}
\usepackage{algorithm}
\usepackage{algpseudocode}
\usepackage{multirow}
\usepackage{array}
\usepackage{booktabs}
\usepackage{subfig}
\usepackage[utf8]{inputenc}
\usepackage{colortbl}
\usepackage{tcolorbox}
\usepackage{multirow}
\usepackage{xcolor}

\definecolor{sea-green}{rgb}{0.4,0.553,0.66}
\definecolor{olive-green}{rgb}{0.561,0.694,0.333}
\definecolor{light-teal}{rgb}{0.443,0.659,0.624}
\definecolor{light-red}{rgb}{0.788,0.4,0.357}

\providecommand{\abs}[1]{\lvert#1\rvert}
\newcolumntype{L}[1]{>{\raggedright\let\newline\\\arraybackslash\hspace{0pt}}m{#1}}
\newcolumntype{C}[1]{>{\centering\let\newline\\\arraybackslash\hspace{0pt}}m{#1}}
\newcolumntype{R}[1]{>{\raggedleft\let\newline\\\arraybackslash\hspace{0pt}}m{#1}}

\newcommand{\gb}{\cellcolor{gray!20}}

\newcommand{\statreport}[1]{\ensuremath{\left[#1\right]}}

\setcopyright{acmcopyright}
\copyrightyear{2021}
\acmYear{2021}

\acmBooktitle{SIGIR '21: The 44th International ACM SIGIR Conference on Research and Development in Information Retrieval, July 11--15, 2021}

\begin{document}
\fancyhead{}

\title{Graph Convolutional Embeddings for Recommender Systems}

\author{Paula Gómez Duran}
\authornote{This work was carried out during an internship in Telefonica Research Barcelona, Spain.}
\affiliation{%
  \institution{University Barcelona, Spain}
}
\email{paula.gomez@ub.edu}

\author{Alexandros Karatzoglou}
\affiliation{%
  \institution{Google Research, London, UK}
}
\email{alexandros.karatzoglou@gmail.com}

\author{Jordi Vitrià}
\affiliation{%
  \institution{University Barcelona, Spain}
}
\email{jordi.vitria@ub.edu}

\author{Xin Xin}
\affiliation{%
  \institution{University of Glasgow}
}
\email{x.xin.1@research.gla.ac.uk}

\author{Ioannis Arapakis}
\affiliation{%
  \institution{Telefonica Research, Barcelona, Spain}
}
\email{ioannis.arapakis@telefonica.com}


\begin{abstract}

Modern recommender systems (RS) work by processing a number of signals that can be inferred from large sets of user-item interaction data. The main signal to analyze stems from the raw matrix that represents interactions. However, we can increase the performance of RS by considering other kinds of signals like the context of interactions, which could be, for example, the time or date of the interaction, the user location, or sequential data corresponding to the historical interactions of the user with the system. These complex, context-based interaction signals are characterized by a rich relational structure that can be represented by a multi-partite graph. 

Graph Convolutional Networks (GCNs) have been used successfully in collaborative filtering with simple user-item interaction data. In this work, we generalize the use of GCNs for N-partide graphs by considering N multiple context dimensions and  propose a simple way for their seamless integration in modern deep learning RS architectures. More specifically, we define a graph convolutional embedding layer for N-partide graphs that processes user-item-context interactions, and constructs node embeddings by leveraging their relational structure. Experiments on several datasets from recommender systems to drug re-purposing show the benefits of the introduced GCN embedding layer by measuring the performance of different context-enriched tasks.

\end{abstract}
%
%


\maketitle

\section{Introduction}

The amount of information available on the internet and beyond far exceeds any human capacity to take account of it. Recommender Systems and in particular collaborative filtering methods have thus become one of the main ways people explore, filter and discover content and information.

One of the undertakings when building an online RS is to use all available data to generate recommendations that are useful and relevant. Data that is collected from online systems on user interactions can include a broader range of information than just the indicators of user-item interactions, such as the time and day of the interaction, the location of the user, the type of the device that acted as the medium for the interaction, the historical interactions of the user during and before that session, as well as the past recommendations the user received during her interactions with the system. This type of meta-data is commonly known as {\em context}~\cite{adomavicius2011context}. 

User, item, and context data are often collected in the form of user-id, item-id and context-id's. These data can be represented in a N-dimensional sparse tensor or, equivalently, as an {\it N-partite graph}, where users, items, and context dimensions define different types of nodes that are connected when there are interactions that involve them. In \autoref{fig:encoded_info}, we show the case of a given user that has interacted with an item under several contexts:

 \begin{figure}[H]
    \centering
    \smallskip
    \includegraphics[trim=0 1cm 0 1cm, width=0.6\linewidth]{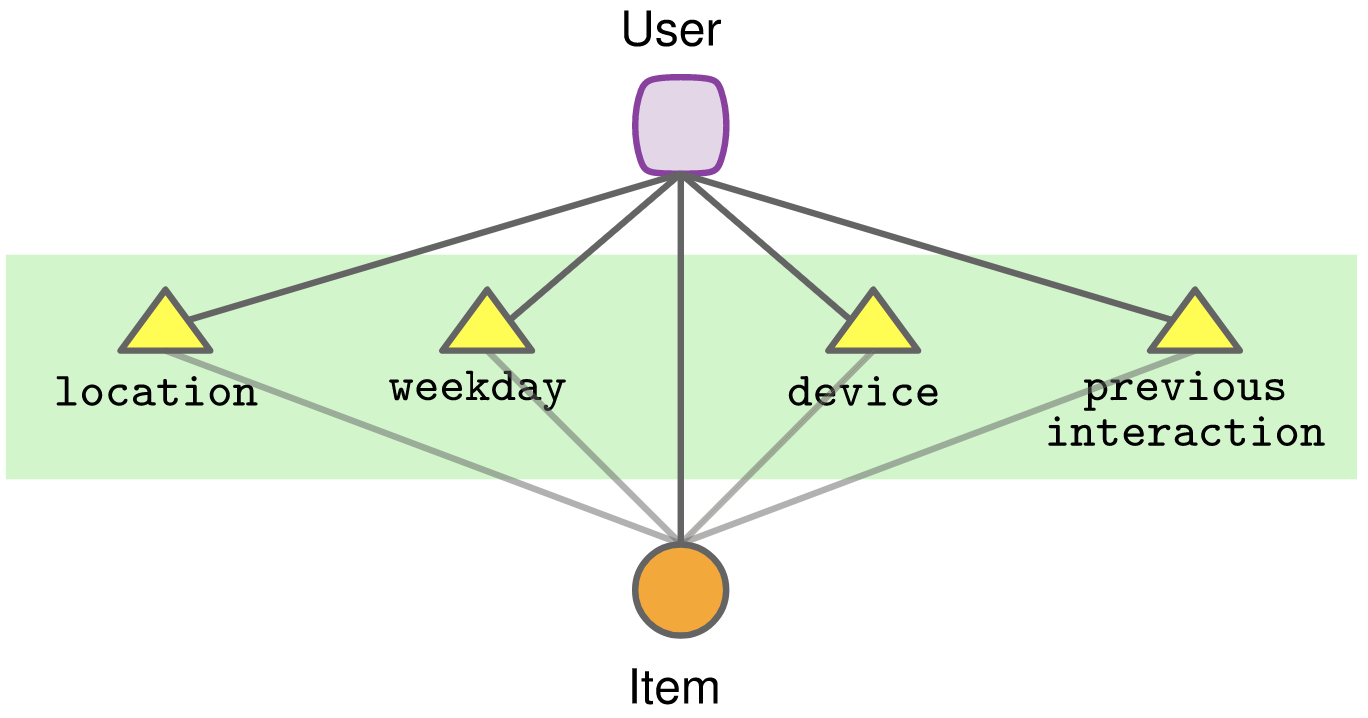}
        \caption{Illustration of a sub-graph from the interaction of a user which interacted with an item under four different context taken into account: weekday, location, device, previous interaction.}
    \label{fig:encoded_info}
\end{figure}

Several methods have been proposed for processing such data. Some of the methods rely on factorizing the user-item-context interactions~\cite{rendle11,rendle2010factorization,karatzoglou10}, while other more recent approaches use deep neural networks to produce powerful representations of the user-item-context interaction data~\cite{guo2017deepfm, nfm, he2017neural}. Typically, Deep Learning (DL) methods compute embeddings at the first layers of the network, which typically consist of standard dense layers~\cite{Barkan16}.   

Embedding vectors, or latent factors, computed by modern collaborative filtering techniques, such as factorization machines or deep factorization machines, rely on a loss function to provide a signal (gradient) about the existence of a user-item interaction. In most of the cases, recommendation is based on pair-wise products between different embedding vectors.  The models used to learn these latent representations are mainly composed of fully connected layers that do not present any interesting inductive bias within.


Graph Convolutional Networks~\cite{Kipf16} have recently gained prominence with applications in several domains, from node classification~\cite{Kipf16} to physical systems simulation~\cite{Sanchez-Gonzalez2020}. 
Learning dense and low-dimensional node representations from graph-structured data has become the keystone in many practical application scenarios, including typical user-item collaborative filtering~\cite{berg2017graph}. Their main advantage is their ability to deal with node information and relational structure at the same time. GCNs learn node representations as the summary of the node's neighborhood, which allows them to effectively encode the graph's (higher-order) topological structure and connectivity information~\cite{Kipf16,bruna2013spectral,henaff2015deep}. Moreover, one of the advantages of GCN is that they seamlessly incorporate graph structure information in the form of the graph adjacency matrix into the model. Hence GCNs do not need to learn that structure and can focus their modeling capacity on learning good node representations. 


Since many types of data come in the form of graphs, or can be converted to graphs, GCNs have proven to be an extremely useful addition to the existing DL modeling approaches. GCNs have been proposed for collaborative filtering algorithms in particular types of recommendation tasks, such as social recommendations~\cite{fan2019graph}, item-to-item recommendations~\cite{ying2018graph}, or rating prediction using user-item ratings ~\cite{berg2017graph}. While these GCN-based recommendation methods demonstrate the potential of GCNs in RS, they are built to address very specific recommendation settings and cannot be easily extended to other architectures.

In this work, we explore how using relational inductive biases within deep learning architectures can facilitate learning about entities in a collaborative context-aware recommendation scenario. To this end, we generalize the use of GCNs in the area of context-aware RS by building a GCN layer that can be used as substitute of a classical embedding layer, allowing its use in practically any recommendation model. This layer can process both implicit and explicit feedback for generalized user-item-context interactions, while also allowing for seamless integration of side information. 
We refer to the proposed layer as Graph Convolutional Embedding (GCE) layer hereinafter.


By applying GCE layers to existing collaborative filtering models, we significantly improve their performance in terms of off-line ranking metrics on several datasets and models. We also report several experiments to understand the causes of this improvement, being the most important conclusion that the use of a GCE, which endows the model with a strong relational bias during the training process, is sufficient to capture a great deal of interactions statistics prior to any learning. 

\section{Related Work}
\label{sec:relatedwork}

Several methods have been proposed to deal with context-aware recommendation data. 
In what follows, we review some related previous research efforts from the perspective of context-aware recommendation and graph convolutions. 

\subsection{Context-aware recommendation}
\label{subsec:matrix-factorization}

The traditional solution for collaborative filtering is Matrix Factorization (MF) ~\cite{matrixFactorization}, even though it presents some clear limitations to model user-item interactions like not being able to incorporate side-features or not dealing properly with sparse data or the cold-start problem \citep{rendle2010factorization}. 

Factorization Machines (FM) \cite{rendle2010factorization,rendle11} generalize MF by making use of inner product to model pairwise latent representations of feature interactions and then formulating their sum to get the predicted scores. They also allow to add side-information of the features and so FM had demonstrable success due to its high performance and low (linear) computational complexity. 

Several efforts have also been made to enhance FM with DL components \cite{he2017neural,guo2017deepfm,xin2019cfm,lian2018xdeepfm}. For example, Neural Factorization Machines (NFM) \cite{nfm} leverages the flexibility of neural networks to replace dot products of matrix factorization and to learn higher-order signals. Deep FM \cite{guo2017deepfm} resembles the original FM and it adds a term of concatenated embeddings. Finally, Neural Collaborative Filtering \cite{he2017neural} borrows ideas from Deep FM, thus combining the FM term with input features instead of their embeddings.


In order to extend the application of MF to context-aware data, Karatzoglou et al ~\cite{karatzoglou10} present an $N$-dimensional tensor factorization method that accounts for context features along with the user-item interactions. In this modelling approach, user-item-context data are represented as tensors $\mathbf{X} \in  \mathbb{R}^{N}$, where $N = m\times n\times c_1\times c_2\dots c_{N-2}$, being  $m$, $n$, $c_1$, \dots, $c_{N-2}$ the dimensions of the user, item and different contexts variables respectively. 
Typically, $\mathbf{X}$ contains non-zero entries at the positions corresponding to a user-item-context interaction and zero (or missing values) otherwise, hence this tensor is extremely sparse. The intuition behind using tensor factorization to solve this problem is that, within a certain context, there are latent features that determine the likelihood of a user-item interaction. 
In the case of FM models, context can be naturally included in FM models as new features. This results in a model that, instead of using a tensor inner product as in \cite{karatzoglou10}, considers the sum of pairwise operations among features.

\subsection{Graph Convolutional Networks}
\label{subsec:gcn}

All methods reported in the previous section \citep{matrixFactorization, rendle2010factorization, nfm, he2017neural} treat input features (user, item or context) as independent elements that are not explicitly related in any way. Thus, they solely rely on the signals of the user-item interactions to capture their latent representations (embeddings). In order to capture embeddings that account for the different relational structures, we propose the use of Graph Convolutional Networks, which allow to capture structure-sensitive latent representations.



Graph Convolutional Networks ~\cite{bruna2013spectral,Duvenaud15,Defferrard16,Kipf16} are a class of neural networks that was introduced for modeling graph-structured data.
The first generation of graph convolution models was motivated from the spectral perspective \citep{bruna2013spectral,henaff2015deep,defferrard2016convolutional}. Based on this body of research, \citet{Kipf16} proposed a new model, known as GCNs, based on the re-normalization trick on the spectral filter and also on graph convolutions for semi-supervised node classification. Due to the success of GCN, both in terms of performance and efficiency, there has been an surge of research on the topic. 
\citet{berg2017graph} proposed to utilize GCN to perform matrix completion on user ratings. 
\citet{graphsage} and \citet{chen2018fastgcn} proposed sampling-based graph convolution. \citet{wu2019SGC} treated the neighborhood aggregation as a pre-computing process and accelerated the training of GCN. \citet{velivckovic2017GAT} proposed to use the attention mechanism, other than topology (degree) information, to learn the node importance when constructing the convolution kernel.

While different versions of GCNs have been proposed in the literature, they all share some basic principles. The assumption of GCNs is that the embedding of a node must result from the aggregation of itself and its neighbors. More specifically, GCNs aim to learn a function of node features on  graph topological data. It is  defined as a function over the graph $\mathcal{G=(V;E)}$, where $\mathcal{V}$ denotes the vertices or nodes of the graph and $\mathcal{E}$ is the edges, which takes the following inputs:
\begin{itemize}
    \item An original feature matrix $Z \in \mathbb{R}^{|\mathcal{V}|\times d}$, where $d$ is the dimension of the original features. Usually, $Z$ is sparse and uses one/multi-hot encoding to represent features.
    \item A representative description of the graph structure in matrix form, which is called adjacency matrix $A \in \mathbb{R}^{|\mathcal{V}|\times |\mathcal{V}|}$. $a_{ij}$ is the $(i,j)$-th entry of $A$ with $a_{ij}=1$ denotes there is an edge between node $i$ and node $j$. Otherwise, $a_{ij}=0$.
\end{itemize}{}

and produces the output, which is a $|\mathcal{V}| \times f$ embedding matrix with $f$ being the output embedding size per node. GCNs can thus be seen as a processing layer in deep networks that takes into account the topological structure of the input graph. Moreover, it is also possible to stack multiple GCN layers.

The layer-wise propagation rule of multi-layer GCN can be formulated as $H^{(l+1)} = \sigma(\hat{D}^{-\frac{1}{2}}\hat{A}\hat{D}^{-\frac{1}{2}}H^{(l)}W^{(l)})$
where $\sigma$ is the activation function, $H^{(l)}$ denotes the hidden representation after $l$-th layer of graph convolution, 
$W^{(l)} \in \mathbb{R}^{d_l \times d_{l+1}}$ is the trainable weight matrix with $d_l$ being the hidden size of the $l$-th layer\footnote{Note that the initial layer of $H$ would be $H^{(0)} = Z$ and $d_0$=$d$. In the final $L$ layer $d_L=f.$};
$\hat{A} = A + I_N$, where $I_N$ is an identity matrix, is the adjacency matrix of the graph $\mathcal{G}$, with added self-connections in order to include the node's own features;  and
$\hat{D}=D+I_N$, where $D$ is the diagonal degree matrix: $D_{ii} = \sum_{j} A_{ij}$. The normalized adjacent matrix $\hat{D}^{-\frac{1}{2}}\hat{A}\hat{D}^{-\frac{1}{2}}$ serves as the keystone to perform neighborhood aggregation between nodes of the graph.

\begin{figure*}
\centering
\includegraphics[trim=0 0.75cm 0 0.05cm, width=0.9\textwidth]{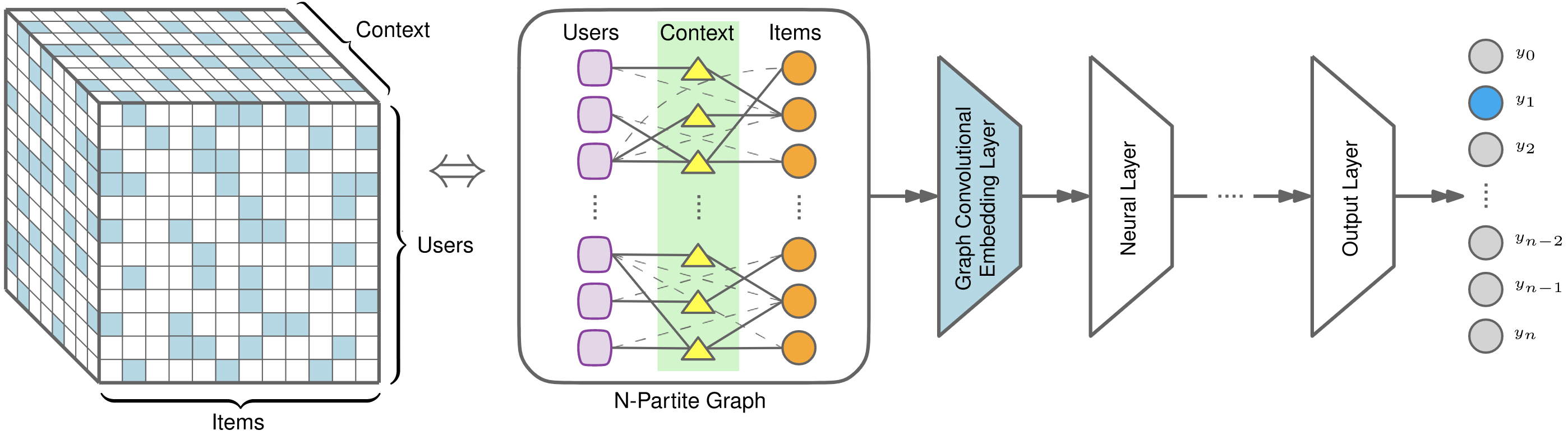}
\caption{
User-item-context interaction tensor $\mathbf{X}$ is equivalent to a three-partite graph $\mathcal{G}$ under the setting of a single context dimension. Edges correspond to interaction events. $\mathcal{G}$ is the input to the GCE layer that generates the node representations that are then used by layers of a recommendation model.}
\label{fig:gce}
\end{figure*}

\paragraph{GCNs in recommender systems}
\label{subsec:gc-mc}

GCN were first introduced to recommendation problems in Graph Convolutional Matrix Completion (GC-MC) work \cite{berg2017graph}, which aimed to perform rating prediction based on user-item interactions. 

GC-MC consists of a graph-convolutional auto-encoder framework which is based on message passing through different kinds of edges (ratings). The rating matrix $R \in \mathbb{R}^{m\times n}$ is used to construct the adjacency matrix of a bipartite graph between user and item nodes. The edges in this graph correspond to the ratings the user have given to the item. Different ratings are captured by different kinds of edges. The learning task is predicting the value of an unobserved entry in $R$ 
and, therefore, can be seen as a link prediction problem.

In GC-MC,   
the latent user and item node representations are based on differentiable message passing on the bipartite interaction graph\footnote{These latent user and item representations are used to reconstruct the rating links through a bi-linear decoder}. The graph convolutional layer performs local operations that take the direct neighbors of a node into account, whereby the same transformation is applied across all locations in the graph. 
A specific transformation is assigned for each rating level, resulting in edge-type specific messages $\mu_{j\rightarrow i,r}$ from items j to users i according to rating $r$. 
The equation is then defined in the following form:
$\mu_{j\rightarrow i,r} = \frac{1}{s_{ij}}W_r z_j$
where $s_{ij}$ is a normalization coefficient that can be either $\abs{\mathcal{N}_i}$ (left normalization) or  $\sqrt{\abs{\mathcal{N}_i}\abs{\mathcal{N}_j}}$ (symmetric normalization), with $\mathcal{N}_i$ denoting the set of neighbors of node $i$; $W_r$ is the trainable weight matrix regarding to the rating $r$ and $z_j$ is the feature vector of node $j$. 

 In~\cite{ying2018graph}, GCNs are used for RS by processing item representations generated by both co-clicking patterns and content-based similarity data. The item similarity graph is processed with graph convolutions to generate recommendations based on the items the users have interacted with. Note that this work considers only the item-similarity graph and not the interactions between users, items and context.
Additional methods utilizing GCNs for collaborative filtering have also been developed, such as those proposed by ~\citet{ngcf}, ~\citet{kgat}, and ~\citet{sunmultigraph}. 

However, all these methods focus on bipartite user-item interactions while our work aims to generate recommendations based on more generalized user-item-context interactions. Thus, in order to use GCN in our work, we will need to extend message passing to multiple dimensions and, then, integrate it in those variations of FM models that naturally allow multiple feature dimensions.








\section{Graph Convolutional Embeddings}
\thispagestyle{fancy}

In this work, we propose a general Graph Convolutional Embedding layer that can be used as a building block in deep neural recommendation methods. We define a graph convolution layer that can model complex user-item-context interactions with an arbitrary number of different contexts. User-item-context data can be represented in a $N$-dimensional tensor $\mathbf{X} \in  \mathbb{R}^{m\times n\times c_1\times c_2\dots c_{N-2}}$, as described in \autoref{subsec:matrix-factorization}. 
These interactions can be modeled as a $N$-partite graph,
where users, items, and context dimensions are different types of nodes that are connected to represent a user-item interaction under a particular context. We show this relationship (for a single context dimension) in \autoref{fig:gce}, the user-item-context data is depicted both as a tensor and a equivalent  $N$-partite graph. 

Graph convolutional embeddings use $N$-partite graphs as input to generate representations for the user, item, and context nodes, which can then be used in downstream recommendation models. Note that, in our model, the existence of context is not necessary and also that GCE works both with implicit and explicit (rating) feedback data, as this only depends on the loss function of the downstream recommendation task.

\subsection{Model}
\label{sec:gfm}
We describe the GCE layer and the node convolutions over the $N$-partite graph. In the GC-MC model \cite{berg2017graph}, the local convolution is seen as a form of message passing from the representations of item $j$ to the representation of user $i$, and vice-versa. We use a different message passing formulation in order to generalize it to context, which can have either one or several dimensions (from $c_1$ to $c_{N-2}$). Here, the messages to each node pass not only from the user (or item) nodes but also from the context nodes that correspond to the user-item interaction. 
That is, when updating a user node $i$, the update includes messages not only from the items that the user interacted with (the item neighborhood of the user) but also 
from the context nodes that have edges with both the user and the item of the update. 

The update rule can be formulated as\footnote{Here the formulation is used for implicit feedback data. For explicit rating data, we just need to add the subscript of rating $r$ to the trainable weight matrices (e.g., $W_u \rightarrow W_{u,r}$).}:
 \begin{equation}
 \label{eq:message-passing-modified_and_extended}
     \mu_{j, e,...,o\rightarrow i} = 
     \frac{1}{s_{ji}}W_u h_j+
     \frac{1}{s_{ei}}W_{c_1} h_e+\dots+
     \frac{1}{s_{oi}}W_{c_{N-2}} h_o
 \end{equation}
where $h_j$ is the input embedding for node $j$. In the first layer, $h_j=z_j$, index $j$ is used to index nodes that correspond to users, $i$ for item nodes, and $e$ for context $c_1$, etc. 
$W_u$ is the trainable weight matrix for the user factors, $W_{c_1}$ for the context $c_1$, etc. $s_{ji}$ is the normalization factor 
$s_{ji}=\sqrt{|\mathcal{N}_j+1||\mathcal{N}_i+1|}$, where $\mathcal{N}_j$ denotes the degree of node $j$. 

Apart from the messages coming from neighbors, the node should also keep its own features. Finally, we apply an element-wise activation function to add non-linearity to our model. As a result, the new embedding for node $i$ is formulated as:
\begin{equation}
\begin{aligned}
 \label{eq:new_node_embeddinng}
     h_i^{(1)}=\sigma(\frac{1}{\mathcal{N}_i+1}W_i h_i+\mu_{j, e,...,o\rightarrow i})
     =\\  \sigma(\frac{1}{\mathcal{N}_i+1}W_i h_i+
      \frac{1}{s_{ji}}W_u h_j +  \dots+
      \frac{1}{s_{oi}}W_{c_{N-2}} h_o)
\end{aligned}
\end{equation}
where the first term denotes the preserved features of node $i$ itself.
If we use a shared weight matrix (i.e., $W_i=W_u=\dots=W_{c_{N-2}}=W$)\footnote{It is a common practical use case to prevent overfitting.}, \autoref{eq:new_node_embeddinng} can be written into a more concise form:
\begin{equation}
 \label{eq:new_node_embeddinng_shared}
     h_i^{(1)}=\sigma((\frac{1}{\mathcal{N}_i+1}h_i+
      \frac{1}{s_{ji}} h_j+\dots+
      \frac{1}{s_{oi}} h_o) W)
 \end{equation}
We can use $H$ to denote the whole input matrix, then \autoref{eq:new_node_embeddinng_shared} can be rewritten into the matrix formulation:
\begin{equation}
	\label{equation:new_emeddinng_matrix}
	H^{(1)}=\sigma(\hat{D}^{-\frac{1}{2}}\hat{A}\hat{D}^{-\frac{1}{2}} HW)
\end{equation}
where $H^{(1)}$ is the new embedding matrix after graph convolution. $\hat{D}^{-\frac{1}{2}}\hat{A}\hat{D}^{-\frac{1}{2}}$ is the normalized adjacent matrix defined on  the the $N$-partite interaction graph (as shown in \autoref{fig:gce}). It conforms with the original definition of GCN \cite{Kipf16}, as discussed in subsection 2.2. 


\subsubsection{Graph Adjacency Matrix Construction}

We construct the adjacency matrix $A$ of the $N$-partite graph based on the user-item-context interaction data. For every interaction, we will have a multi-field categorical vector 
which contains the indexes of the \textsf{ <userID, itemID, contextID\textsubscript{1}, contextID\textsubscript{2}, ..., contextID\textsubscript{N-2}> } of the interaction. Note that we re-index all the nodes in order to provide a unique identifier for all. 
As a result, the total number of nodes in the constructed graph is $m+n+c_1+\dots+c_{N-2}$ (i.e., $|\mathcal{V}|=m+n+c_1+\dots+c_{N-2}$). 
\vspace{-1em}

\begin{figure}[H]
    \centering
    \smallskip
    \includegraphics[width=0.75\linewidth]{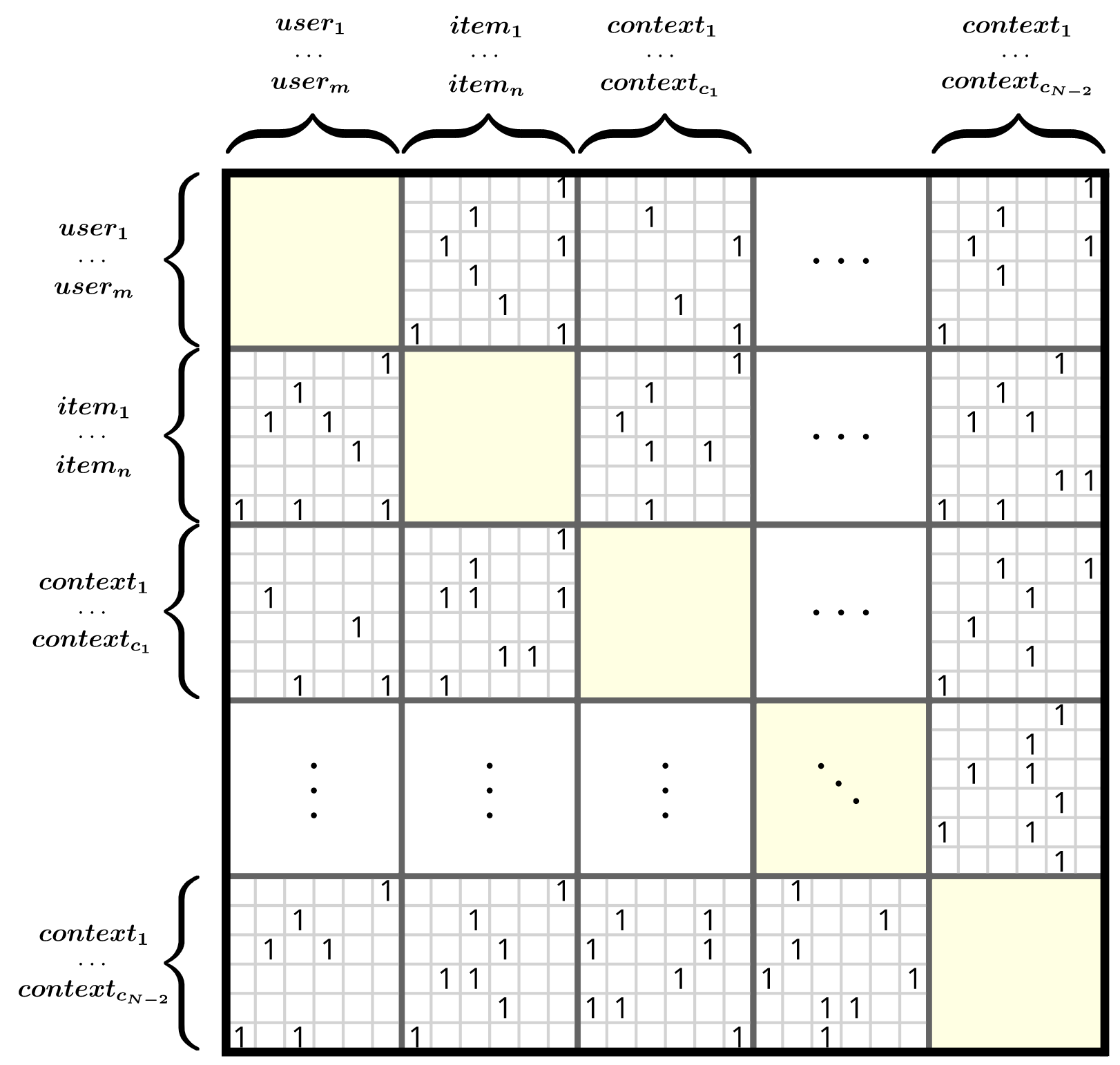}
    \vspace{-0.5em}
        \caption{Schema providing an intuition of how matrix $A$ is built.}
    \label{fig:adjacency_mx}
\end{figure}
\vspace{-1.5em}
An illustration of the adjacency matrix construction is shown in \autoref{fig:adjacency_mx}. Note that $A$ is built just with the training set, and naturally has a block-diagonal structure. Note that user-items, user-context and item-context are connected with bidirectional links while user-user, item-item and context-context nodes are not linked, depicting that context can influence the user preference or how an item is perceived. 

\subsection{GCE for Factorization Machines}

In this section we describe the use of GCE with FM \cite{rendle11}, namely GCE-FM. Note that the GCE layer can be used essentially with any recommendation model that uses and computes embeddings through gradient descent, such as 
MF~\cite{baltrunas2011matrix}, NFM~\cite{nfm}, NCF~\cite{he2017neural}, etc. (see \autoref{subsec:baselines} for a detailed description of some of those models).

The combination of GCE with the FM model is relatively straightforward. While the standard FM models pairwise interactions between different context variables through inner product, the result is linear to the embedding factors.
When combined with GCE, the result remains linear with respect to the output of the GCE layer. 
Note that the GCE layer is fully differentiable, hence the model parameters can be easily computed with back-propagation and stochastic gradient descent. 
The embeddings computed by the GCE layer already capture high-order interaction signals between the different nodes of the user-item-context graph; hence FM, in effect, acts as an additional layer of interactions computation. 

An entry (interaction record) in the original input tensor $\mathbf{X}$ can be represented as a sparse categorical vector $x \in \mathbb{R}^{|\mathcal{V}|}$. Then, the predicted score based on the input $x$ can be formulated as:
\begin{equation}
\label{eq:GEM}
    \hat{y}(x) = w_0 + \sum_{p=1}^{|\mathcal{V}|}w_p x_p + \sum_{p=1}^{|\mathcal{V}|}\sum_{q=p+1}^{|\mathcal{V}|} x_p x_q \langle \boldsymbol{g}(x_p),\boldsymbol{g}(x_q) \rangle
\end{equation}{} 

where $x_p$ denotes the $p$-th element in $x$ (corresponding to a context variable $p$),  $w_0$ is the global bias, $w_p$ models the bias of $p$. $\boldsymbol{g}(x_p)$ is the output GCE embedding corresponding to the context variable $p$, and $\langle \cdot \rangle$ denotes inner product of two vectors.
From this perspective, the original FM would be a special case of GCE-FM, when $\boldsymbol{g}(x_p)$ is simplified to an embedding table lookup operation without graph convolution. Thus, for any other RS-models we can compute it's GCE-version like: GCE-NFM, GCE-WD, GCE-DFM... by just changing the way of capturing their latent representations.


 


\subsection{Training}

After combing GCE with the downstream recommendation model, we train the whole parameters in an end-to-end fashion. For top-$K$ recommendation, we use the BPR loss \cite{rendle2012bpr} as the training function:


\begin{equation}
\label{eq:bce_loss_with_logits}
  L = -\sum{\log(\sigma((S(u;i;c) - S(u;j;c)))}
\end{equation}
where $S(u;i;c)$ is the score prediction of a positive interaction between user $u$ and an item $i$ in a particular context $c$, and $S(u;j;c)$ is the score prediction of the corresponding negative sample (same $u$ and $c$ but sampled item $j$ denoting non-interaction between $u$-$j$ in a given $c$).

\subsection{Time complexity}
Since GCE can be seamlessly combined with downstream recommendation models, we only need to consider the additional cost introduced by the graph convolution operation.
Consider a GCE layer as shown in \autoref{equation:new_emeddinng_matrix}. As the normalized adjacent matrix can be pre-computed, the additional complexity only comes from two matrix multiplication operations. 

The first multiplication is between the normalized adjacent matrix (i.e., $\hat{D}^{-\frac{1}{2}}\hat{A}\hat{D}^{-\frac{1}{2}}$) and the input matrix. However, due to the sparse structure of the adjacent matrix, the complexity is only related to the number of non-zero entries in the matrix. The second matrix multiplication operation is a standard matrix multiplication with a complexity linear to the number of nodes $|V|$. As a result, the total computational cost of a GCE layer is linear with respect to the number of nodes and edges of the constructed graph~$O(\mathcal{|V||E|}d^2)$.
\section{Experiments}
\label{sec:experiments}
\thispagestyle{fancy}

In this section we answer several experimental questions: 
\begin{enumerate}
    \item  Does the performance of state-of-the-art context-aware recommendation models increase in a consistent way when using GCE?
    \item What part of the performance increment is due to the relational inductive bias that is encoded within GCE?
    \item How can we measure the utility of different relational structures for recommending items? 
\end{enumerate}

First, we evaluate the proposed GCE by combining it with three state-of-the-art recommendation models, which can be trained by back-propagation and which can naturally be extended to do context-aware recommendations. 
We compare the performance of every model in two different situations: by using a GCE layer and by using a traditional embedding layer. We do so by training all models end-to-end on the same dataset and also using the same loss function and criteria to decide all meta-parameters. 

Additionally, we evaluate the case of incorporating side-infor\-mation (if the dataset provides it) in some of the nodes in order to show how GCE can easily incorporate any kind of information, either as nodes or as information inside each node. We evaluate for each dataset whether GCE layer helps to improve the performance of the model or not.
For example, taking FM as a baseline model we will compare the three following models, all trained end-to-end:

\begin{itemize}
    \item \textit{FM}: the selected baseline model (in this case FM)
    \item \textit{FM-GCE}: baseline model using GCE layer instead of usual embeddings
    \item \textit{FM-GCE-SINFO}: baseline model using GCE layer with side-information (represented as input features) for some nodes
\end{itemize}

\noindent We perform the comparisons on 
four benchmark datasets:  MovieLens-100k\cite{harper2015movielens}, LastFM-1K\cite{zangerle2014nowplaying}, User-Book Interactions\cite{wan2019fine, wan2018item} and Drug-Disease Interactions\cite{Luo2017}. On the one hand, the first three datasets consist of user-item interactions on a diverse set of items: movies, songs and books, respectively. 
In addition, they all hold implicit context information that can be inferred from the data; for example,  we can determine the order of the item(s) that the user interacted with by sorting all items of a given user according to the interaction's timestamp. 
On the other hand, the last dataset\footnote{Available at https://github.com/luoyunan/DTINet/} consists of interactions between drugs ("users") and diseases ("items"), originally used for drug-target interaction prediction and drug re-purposing \cite{Luo2017}. The dataset includes protein interaction data for both drugs and diseases, which we use as context by doing the intersection of those which were present when a drug-disease interaction occurred. In addition, it also provides side-effect information that is  associated to drugs, which we can use as side-information.  

Furthermore, in our experiments, we focus on the top-$K$ recommendation task based on implicit feedback data, which is a more common scenario compared with rating predictions.  
As a result, we transform the dataset that contain ratings to binary user-item interactions and we also apply some pre-processing for each of the datasets as discussed in \ref{subsec:performance}. We end up with the following statistics for each of the datasets after applying pre-processing\footnote{Statistics reported after applying all data transformations needed.}:

{\def\arraystretch{0.95}
\begin{table}[H]
\centering
{\begin{tabular}{L{2cm}C{1.25cm}C{1.25cm}C{2cm}}
    \toprule
    \textbf{dataset} & \textbf{\#users} & \textbf{\#items} & \textbf{\#interactions} \\
    \midrule
    \texttt{ML-100K} & 943 & 1.682 & 100.000  \\
    \texttt{Books} & 3.909 & 1.412 & 172.124  \\
    \texttt{LastFM} & 709 & 5.000 & 468.740  \\
    \texttt{Drug-Disease} & 549 & 2.575 & 127.406  \\
    \arrayrulecolor{black}\bottomrule
\end{tabular}}

\caption{Dataset statistics in terms of number of users, number of items and number of interactions among them. }
\label{table:data-statistics}
\end{table}
}

\vspace{-3em}

\subsection{Experimental setup}
\subsubsection{Baselines}
\label{subsec:baselines}


We implemented the following baseline models using Pytorch\cite{ketkar2017introduction}. For each of these baselines, we tested the performance of the three variants as discussed before\footnote{Third variant will be done just if dataset provides side-information.}. Note that all baselines need to be and thus are context-aware i.e. they do model user-item-context interactions. Moreover all baselines are trained and tested on the same data, being tuned following the same criteria and using the same loss function. 

\begin{itemize}
    \item \textbf{Matrix Factorization (MF)} \cite{baltrunas2011matrix}:  This is the classic matrix factorization method. We extended it to tensor factorization (decomposition in n matrices instead of just two) in order to allow using context.
    
    \item \textbf{Factorization Machines (FM)} \cite{rendle2010factorization}:  This is the original FM. It is a strong baseline that computes the pair-wise operation between features instead of computing the dot product across all dimensions.
    


    \item \textbf{Neural Collaborative Filtering (NCF)} \cite{he2017neural}: This baseline resembles the original FM, but adds another term of concatenated input vectors that are fed into a MLP.
    
\end{itemize}

\subsubsection{Parameter settings}

To ensure a fair comparison of the models' performance, we train all of them by optimizing the BPR loss\cite{rendle2012bpr} with the Adam optimizer\cite{kingma2014adam}. We use Bayesian Optimization\cite{bergstra2015hyperopt} strategy in order to tune the hyperparameters and follow the same criteria in all cadses. Thus, we determine the best hyperparameters for each of the models by tuning the learning rate on the range [0.0001, 0.0005, 0.001, 0.005, 0.01]; batch size on the range [256, 512, 1024, 2048], and dropout on range [0, 0.15, 0.5]. The embedding size is set to 64 for all models\footnote{In this paper, we report the results when using one GCE layer because it will not introduce additional trainable parameters compared with the commonly used embedding table.}. We run all the experiments for a maximum of 150 epochs and perform early stopping when the evaluation metrics defined in \autoref{sec:evaluation} stops improving for more than 5 consecutive epochs.

\subsubsection{Evaluation protocol}
\label{sec:evaluation}

The evaluation of all models follows the standard offline top-$K$ evaluation, where the target is to generate a ranked list (according to the predicted scores) of $K$ items that a user is most likely to interact with. We use the \textit{leave-one-out} strategy for splitting the dataset to train validation and test sets. This strategy has been widely used in literature \cite{he2018nais, yuan2016lambdafm, he2017neural}. We held out the last two interactions of the users for the validation and test set, if timestamp is available. We, thus, require users to have interacted with at least three items so that we can -at a minimum- allocate one observation for training, one for validation and one for testing. The datasets have been pre-processed based on this constraint (we provide a more detailed description of the pre-processing in \autoref{subsec:performance}).

We report the Hit Ratio (HR) and Normalized Discounted Cumulative Gain (NDCG) metrics on the test set by ranking among all items for evaluating the recommendation quality of the models:

\begin{itemize}
    \item \textbf{HR@K:} A recall-based metric, measuring whether the test item is in the top-$K$ positions of the recommendation list. 
    \item \textbf{NDCG@K:} It is a rank-sensitive measure of quality, which provides information on the ranking position of the ground truth sample; higher scores are assigned to the top-ranked items and can be regarded as a weighted version of HR@K.

\end{itemize}

When evaluating the performance, the models give predicted scores (logits) of each item for a given user and context variables in the validation or test set. All items are sorted in the decreasing order of their scores and top-K items are returned as recommendation list. If the ground truth item is presented in the recommendation list, HR would be 1 for this user, otherwise, it would be 0. The final HR is reported after averaging for all users and an analogous calculation is done for NDCG. All experiments are conducted with 10 different random seeds and the average scores are reported. 

Finally, to investigate further the performance differences across the different model variants, we run Friedman's ANOVA as an omnibus test and, if the result is statistically significant, we use the Wilcoxon signed-rank test for pairwise comparisons (for which we correct the level of significance to control the false discovery rate by using the Benjamini-Hochberg correction~\cite{Benjamini1995,Benjamini2000}). 
Both Friedman and Wilcoxon tests are non-parametric, which means they do not make any assumptions about the data distribution and are therefore appropriate for our experimental setup~\cite{10.5555/1248547.1248548}. We highlight the statistically significant differences between the baselines and their -GCN and -SINFO variants in Tables 2-4 and 6.

\subsection{Performance}
\label{subsec:performance}


We give a brief description of each dataset and the respective context, as well as a description of the pre-processing steps that we applied. We train the standard baseline model, as well as its corresponding GCE variant, without and with side-information (if data provides it), for every dataset and baseline. Please, note that the three versions of each model are trained end-to-end by using context information into account (if using GCE in a $N$-partite graph).


\subsection*{ML-100k dataset} 
The MovieLens 100k dataset\footnote{\url{https://grouplens.org/datasets/movielens}} contains 100.000 ratings from 943 users on 1.682 movies. 
The ratings are provided in the form of \textsf{<userID, itemID, rating, timestamp>} tuples and each user has a minimum of 20 ratings. For this dataset, we used  the last interacted movie of each user as the context. The context is the previous movie that a user selected before the current item interaction. 

As they provide the gender of each of the movies, they are used as side-information to train each baseline one step further when applying GCE layer in \textit{BASELINE-GCE-SI} experiment\footnote{Note that $SI$ refers to Side-Information.}.

{\def\arraystretch{0.8}
\begin{table}[H]
\centering
{\begin{tabular}{L{2cm}C{1.15cm}C{1.15cm}C{1.15cm}C{1.15cm}}
    \toprule
    \multicolumn{5}{c}{\texttt{ML-100k} \statreport{\texttt{last-clicked-item}}} \\ 
    \midrule
    & \multicolumn{2}{c}{\textbf{Top-10}} & \multicolumn{2}{c}{\textbf{Top-20}} \\
    \cmidrule(l{1pt}r{1pt}){2-3}\cmidrule(l{1pt}r{1pt}){4-5}
    & \textbf{HR} & \textbf{NDCG} & \textbf{HR} & \textbf{NDCG} \\
    \midrule
    \gb\texttt{MF} & \gb 0.135 & \gb 0.071 & \gb 0.210 & \gb 0.089 \\
    \texttt{MF-GCE} & \textbf{0.222}\rlap{$^{**}$} & \textbf{0.119}\rlap{$^{**}$} & \textbf{0.329}\rlap{$^{**}$} & \textbf{0.145}\rlap{$^{**}$} \\
    \texttt{MF-GCE-SI} & 0.214\rlap{$^{**}$} & 0.114\rlap{$^{**}$} & 0.327\rlap{$^{**}$} & 0.143\rlap{$^{**}$} \\
    
    \arrayrulecolor{gray!50!}\midrule
    \gb\texttt{FM} & \gb 0.186 & \gb 0.099 & \gb 0.286 & \gb 0.124 \\
    \texttt{FM-GCE} & 0.220\rlap{$^{**}$} & 0.115\rlap{$^{**}$} &\textbf{ 0.329}\rlap{$^{**}$} & \textbf{0.145}\rlap{$^{**}$} \\
    \texttt{FM-GCE-SI} & \textbf{0.222}\rlap{$^{**}$} & \textbf{0.118}\rlap{$^{**}$} & 0.327\rlap{$^{**}$} & 0.144\rlap{$^{**}$} \\
    
    \arrayrulecolor{gray!50!}\midrule
    \gb\texttt{NCF} & \gb 0.186 & \gb 0.100 & \gb 0.279 & \gb 0.124 \\
    \texttt{NCF-GCE} &\textbf{ 0.212}\rlap{$^{**}$} & \textbf{0.113}\rlap{$^{**}$} & 0.305\rlap{$^{**}$} & 0.136\rlap{$^{**}$} \\
    \texttt{NCF-GCE-SI} & 0.211\rlap{$^{**}$} & 0.112\rlap{$^{**}$} & \textbf{0.310}\rlap{$^{**}$} &\textbf{0.137}\rlap{$^{**}$} \\

    \arrayrulecolor{black}\bottomrule
\end{tabular}}
\caption{Performance comparison for all baselines and their GCE variants, without and with movie's gender as item's side-information, for the ML-100k dataset. Sig. levels (two tails, corrected for mult. comparisons): $^{*} p<.01$; $^{**} p<.001$.}
\label{table:ml-results}
\end{table}
}

\vspace{-2em}

As mentioned earlier, we converted the ratings to binary data to use them as implicit feedback. This is done by taking all interactions as positive samples and performed negative sampling in every update step to generate non-interaction samples. No other pre-processing is applied. 

\autoref{table:ml-results} shows the results of the selected baselines when generating top-10 and top-20 recommendation. 
We observe that all models performed significantly better when using GCE and that the best results are achieved by the the GCE layer when also adding side-information. However, adding side-information in MF-GCE does not seem to help. Although, it should be noted that MF-GCE results achieve either the same or the best performance among the nine (9) model variants on this data. This might strengthen the statements from \cite{rendle2020neural}, who states that, with a proper hyperparameter selection, a simple dot product substantially outperforms more expressive models.

\subsection*{LastFM-1k dataset}

The LastFM-1K\footnote{\url{http://ocelma.net/MusicRecommendationDataset/lastfm-1K.html}} is a subset of LastFM\footnote{\url{https://www.last.fm}} which offers a user full listening history containing the whole listening habits (till May, 5th 2009) for nearly 1,000 users. This dataset contains \textsf{<user, timestamp, artist, song>} tuples collected from Last.fm API 
. In our experiments, we used the records of the last year (2009). Furthremore, we select those users with at least 3 interactions and the 5,000 most frequent songs. We used the timestamp to select the last song that the user interacted with, before each listening event, as the context, and the artist as side-information of the user. 

\autoref{table:music-results} shows the results of the selected methods without and with the GCE layer in two different scenarios: not using or using artists as item's side-information. We can see how GCE layer consistently improve the results in this data as well. However, we also note how adding side-information does not lead to better results when the model is expressive enough. This might be because side-information does not introduce any new features that allow the model to learn better representations and thus, using artist as side-information may not be helpful on the depicted scenario.
\vspace{-1em}

{\def\arraystretch{0.8}
\begin{table}[H]
\centering
{\begin{tabular}{L{2cm}C{1.15cm}C{1.15cm}C{1.15cm}C{1.15cm}}
    \toprule
    \multicolumn{5}{c}{\texttt{LastFM} \statreport{\texttt{last-clicked-item}}} \\ 
    \midrule
    & \multicolumn{2}{c}{\textbf{Top-10}} & \multicolumn{2}{c}{\textbf{Top-20}} \\
    \cmidrule(l{1pt}r{1pt}){2-3}\cmidrule(l{1pt}r{1pt}){4-5}
    & \textbf{HR} & \textbf{NDCG} & \textbf{HR} & \textbf{NDCG} \\
    \midrule
    \gb\texttt{MF} & \gb 0.328 & \gb 0.258 & \gb 0.365 & \gb 0.269 \\
    \texttt{MF-GCE} & 0.434\rlap{$^{**}$} & 0.331\rlap{$^{**}$} & \textbf{0.495}\rlap{$^{**}$} & 0.342\rlap{$^{**}$} \\
    \texttt{MF-GCE-SI} & \textbf{0.435}\rlap{$^{**}$} & \textbf{0.333}\rlap{$^{**}$} & 0.492\rlap{$^{**}$} & \textbf{0.347}\rlap{$^{**}$} \\
    
    \arrayrulecolor{gray!50!}\midrule
    \gb\texttt{FM} & \gb 0.458 & \gb 0.358 & \gb 0.513 & \gb 0.367 \\
    \texttt{FM-GCE} & \textbf{0.472}\rlap{$^{*}$} & \textbf{0.368} & \textbf{0.524}\rlap{$^{*}$} & \textbf{0.380}\rlap{$^{**}$} \\
    \texttt{FM-GCE-SI} & 0.471\rlap{$^{*}$} & 0.365 & 0.523\rlap{$^{*}$} & 0.378\rlap{$^{*}$} \\

    
     \arrayrulecolor{gray!50!}\midrule
    \gb\texttt{NCF} & \gb 0.448 & \gb 0.350 & \gb 0.493 & \gb 0.360 \\
    \texttt{NCF-GCE} & \textbf{0.508}\rlap{$^{**}$} & \textbf{0.424}\rlap{$^{**}$} & \textbf{0.553}\rlap{$^{**}$} & \textbf{0.434}\rlap{$^{**}$} \\
    \texttt{NCF-GCE-SI} &  0.489\rlap{$^{*}$} & 0.416\rlap{$^{**}$} & 0.540\rlap{$^{**}$} & 0.428\rlap{$^{**}$} \\

    \arrayrulecolor{black}\bottomrule
\end{tabular}}
\caption{Performance comparison for all baselines and their GCE variants, without and with song's artist as item's side-information, for the LastFM dataset. Sig. levels (two tails, corrected for mult. comparisons): $^{*} p<.01$; $^{**} p<.001$.}
\label{table:music-results}
\end{table}
}
\vspace{-2em}


\subsection*{User-Books Interaction dataset} 

The User-Books Interaction dataset\footnote{\url{https://sites.google.com/eng.ucsd.edu/ucsdbookgraph/shelves}} is a set of user reading history. The extended version of this dataset (aprox. 64GB of size) includes, in addition, the associated timestamps. In our experiments, 
we randomly sampled approximately 1M interactions from 6,000 different users and, in addition, ranked the items by their frequency to acquire the top 5000 read books. Then, we filter the users and force them to have, at least, 20 interactions per user, as done in the Movielens dataset. The final dataset included 172,124 interactions from 3,909 users on 1,412 books. The context used for this dataset is the last book that the user read before the current interaction and it does not provide any side-information. 

{\def\arraystretch{0.8}
\begin{table}[H]
\centering
{\begin{tabular}{L{2cm}C{1.15cm}C{1.15cm}C{1.15cm}C{1.15cm}}
    \toprule
    \multicolumn{5}{c}{\texttt{Books} \statreport{\texttt{last-clicked-item}}} \\ 
    \midrule
    & \multicolumn{2}{c}{\textbf{Top-10}} & \multicolumn{2}{c}{\textbf{Top-20}} \\
    \cmidrule(l{1pt}r{1pt}){2-3}\cmidrule(l{1pt}r{1pt}){4-5}
    & \textbf{HR} & \textbf{NDCG} & \textbf{HR} & \textbf{NDCG} \\
    \midrule
    \gb\texttt{MF} & \gb 0.061 & \gb 0.039 & \gb 0.085 & \gb 0.045 \\
    \texttt{MF-GCE} & \textbf{0.155}\rlap{$^{**}$} & \textbf{0.093}\rlap{$^{**}$} & \textbf{0.217}\rlap{$^{**}$} & \textbf{0.109}\rlap{$^{**}$} \\
    
    \arrayrulecolor{gray!50!}\midrule
    \gb\texttt{FM} & \gb 0.140 & \gb 0.089 & \gb 0.192 & \gb 0.101 \\
    \texttt{FM-GCE} & \textbf{0.160}\rlap{$^{**}$} & \textbf{0.099}\rlap{$^{**}$} & \textbf{0.214}\rlap{$^{**}$} & \textbf{0.113}\rlap{$^{**}$} \\
    
    \arrayrulecolor{gray!50!}\midrule
    \gb\texttt{NCF} & \gb 0.138 & \gb 0.088 & \gb 0.192 & \gb 0.102 \\
    \texttt{NCF-GCE} & \textbf{0.150}\rlap{$^{**}$} & \textbf{0.090} & \textbf{0.213}\rlap{$^{**}$} & \textbf{0.106}\rlap{$^{*}$} \\
    \arrayrulecolor{black}\bottomrule
\end{tabular}}
\caption{Performance comparison for all baselines and their GCE variants, with $k=1$ (default) and $k=2$ for the User-Books Interaction dataset. Sig. levels (two tails, corrected for mult. comparisons): $^{*} p<.01$; $^{**} p<.001$.}
\label{table:books-results}
\end{table}
}
\vspace{-2em}

\autoref{table:books-results} summarises the results of the experiments on the User-Books Interaction dataset. All methods perform much better with the GCE layers, but, analogous to previous results, the improvement on the MF is more pronounced for being the one capturing less high-order interaction information.

{\def\arraystretch{0.8}
\begin{table*}[t!]
\centering
{\begin{tabular*}{\textwidth}{C{0.3cm}C{0.85cm}C{1.25cm}C{1.15cm}C{0.85cm}C{1.25cm}C{1.15cm}C{0.85cm}C{1.25cm}C{1.15cm}C{0.85cm}C{1.25cm}C{1.15cm}}
    \toprule
    & \multicolumn{6}{c}{\texttt{(a) w/o Top-k popular Items}} & \multicolumn{6}{c}{\texttt{(b) w/o Top-k popular Users}} \\ 
    \cmidrule(l{1pt}r{1pt}){1-7}\cmidrule(l{1pt}r{1pt}){8-13}
    & \multicolumn{3}{c}{\textbf{ML-100K}} & \multicolumn{3}{c}{\textbf{User-Books}} & \multicolumn{3}{c}{\textbf{ML-100K}} & \multicolumn{3}{c}{\textbf{User-Books}} \\
    \cmidrule(l{1pt}r{1pt}){2-4}\cmidrule(l{1pt}r{1pt}){5-7}\cmidrule(l{1pt}r{1pt}){8-10}\cmidrule(l{1pt}r{1pt}){11-13}
    \textbf{k} & \textbf{FM} & \textbf{FM-GCE} & \textbf{Improv.} & \textbf{FM} & \textbf{FM-GCE} & \textbf{Improv.} & \textbf{FM} & \textbf{FM-GCE} & \textbf{Improv.} & \textbf{FM} & \textbf{FM-GCE} & \textbf{Improv.} \\
    \cmidrule(l{1pt}r{1pt}){1-7}\cmidrule(l{1pt}r{1pt}){8-13}
    0 & 0.0726  & \textbf{0.0815} & 12.25\% & 0.0567 & \textbf{0.0665}  & 17.28\% & 0.0726  & \textbf{0.0815} & 12.25\% & 0.0567 & \textbf{0.0665} & 17.28\% \\
    10 & 0.0067 & \textbf{0.0078} & 16.41\% & 0.0045 & \textbf{0.0052}  & 15.55\% & 0.0731  & \textbf{0.0789} & 10.79\% & 0.0572 & \textbf{0.0667}  & 16.61\% \\
    25 & 0.0063  & \textbf{0.0074} & 17.46\% & 0.0039 & \textbf{0.0049}  & 25.64\% & 0.0774 & \textbf{0.0863} & 11.49\% & 0.0594 & \textbf{0.0683}  & 14.99\% \\
    50 & 0.0053  & \textbf{0.0072} & 35.84\% & 0.0035 & \textbf{0.0040}  & 14.28\% & 0.0859 & \textbf{0.0911} & 6.05\% & 0.0600 & \textbf{0.0689}  & 14.83\% \\
    \arrayrulecolor{black}\bottomrule
\end{tabular*}}
\caption{Recommendation Performance in NDCG@10 for FM baseline on ML-100k and User-Books Interaction dataset after removing the Top-k popular Items (a) or after removing   the Top-k popular Users (b) in the Test Stage}
\label{table:analysis}
\vspace{-3mm}
\end{table*}
}

\subsection*{Drug–Disease Interaction dataset}

The Drug-Disease Interaction dataset\footnote{\url{https://github.com/luoyunan/DTINet/tree/master/data}} is available for drug-target interaction prediction and drug re-purposing\cite{Luo2017}. 

In this experiments, we use the drug-disease associations as user-item relationships. Hence we are trying to predict good new disease targets for the drugs to be re-purposed. For both drugs and diseases, their respective proteins associations are provided, we compute the intersection of proteins on a drug-disease interaction and use it as context variable. Hence we construct the 3-partite drug-disease-protein graph. Given a drug and a set of associated proteins, we try to predict a ranking of diseases for which the drug can be useful. Additionally, we use the side-effects provided for drugs as side-information on the last experiment to see whether it can increase the performance of a model in this case.

{\def\arraystretch{0.8}
\begin{table}[H]
\centering
{\begin{tabular}{L{2cm}C{1.15cm}C{1.15cm}C{1.15cm}C{1.15cm}}
    \toprule
    \multicolumn{5}{c}{\texttt{Drug–Disease} \statreport{\texttt{proteins-intersection}}} \\ 
    \midrule
    & \multicolumn{2}{c}{\textbf{Top-10}} & \multicolumn{2}{c}{\textbf{Top-20}} \\
    \cmidrule(l{1pt}r{1pt}){2-3}\cmidrule(l{1pt}r{1pt}){4-5}
    & \textbf{HR} & \textbf{NDCG} & \textbf{HR} & \textbf{NDCG} \\
    \midrule
    \gb\texttt{MF} & \gb 0.186 & \gb 0.119 & \gb 0.251 & \gb 0.135 \\
    \texttt{MF-GCE} & \textbf{0.270}\rlap{$^{**}$} & \textbf{0.166}\rlap{$^{**}$} & \textbf{0.354}\rlap{$^{**}$} & \textbf{0.185}\rlap{$^{**}$} \\
    \texttt{MF-GCE-SI} & 0.263\rlap{$^{**}$} & 0.160\rlap{$^{**}$} & 0.351\rlap{$^{**}$} & 0.182\rlap{$^{**}$} \\
    
    \arrayrulecolor{gray!50!}\midrule
    \gb\texttt{FM} & \gb 0.307 & \gb 0.196 & \gb 0.399 & \gb 0.218 \\
    \texttt{FM-GCE} & 0.345\rlap{$^{**}$} & 0.225\rlap{$^{**}$} & \textbf{0.453}\rlap{$^{**}$} & \textbf{0.252}\rlap{$^{**}$} \\
    \texttt{FM-GCE-SI} & \textbf{0.356}\rlap{$^{**}$} & \textbf{0.227}\rlap{$^{**}$} & 0.450\rlap{$^{**}$} & 0.251\rlap{$^{**}$} \\
    
    \arrayrulecolor{gray!50!}\midrule
    \gb\texttt{NCF} & \gb 0.309 & \gb 0.200 & \gb 0.398 & \gb  0.221 \\
    \texttt{NCF-GCE} & \textbf{0.364}\rlap{$^{**}$} & \textbf{0.236}\rlap{$^{**}$} & 0.451\rlap{$^{**}$} & \textbf{0.256}\rlap{$^{**}$} \\
    \texttt{NCF-GCE-SI} & 0.362\rlap{$^{**}$} & 0.232\rlap{$^{**}$} & \textbf{0.452}\rlap{$^{**}$} &0.255\rlap{$^{**}$} \\ 
    \arrayrulecolor{black}\bottomrule
\end{tabular}}
\caption{Performance comparison for all baselines and their GCE variants, without and with side-effects as drug's side-information, for the Drug-Disease dataset. Sig. levels (two tails, corrected for mult. comparisons): $^{*} p<.01$; $^{**} p<.001$.}
\label{table:drugs-results}
\end{table}
}
\vspace{-2em}

As we can observe in \autoref{table:drugs-results}, for the two most expressive models (FM, NCF), the more data we incorporate the more improvement we get. This is because, in this scenario, data fits perfectly a graph-structure. Therefore, adding the proteins intersection of a given drug and disease helps by itself the learning of the model. Additionally, incorporating this information in graph-based structure leads for even more improvements. Analogously, the side information we present here is actually pharmacological or phenotypic information (side-effects) that is specific and directly related with each of the drugs which it is associated with. This allow to show in full essence that all the functionalities of using GCE layer work much better when they is based on data that perfectly has graph shape structure. 

\begin{figure}[!ht]
\centering
\includegraphics[trim=0 0.05cm 0 0.05cm, width=0.95\linewidth]{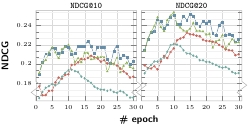}
\caption{Comparison for FM baseline in Drug-Disease Interaction dataset: drug-disease interaction with FM $(\textcolor{light-teal}{\blacklozenge})$, drug-disease-protein interaction with FM $(\textcolor{light-red}{\bullet})$, drug-disease-protein interaction with FM-GCE $(\textcolor{olive-green}{\blacktriangle})$, and drug-disease-protein interaction with FM-GCE-SI, using side-effect as drug's features $(\textcolor{sea-green}{\blacksquare})$.}
\label{fig-graph} 
\end{figure}

In \autoref{fig-graph} we show the performance in terms of NDCG@10 and NDCG@20 of FM baseline in four possible scenarios:

\begin{enumerate}
    \item Drug-disease interactions using FM baseline\footnote{It is analogous to a non-context scenario, like user-item problems.}.
    \item Drug-proteins-disease\footnote{Using protein-intersection as context, as described previously.}, using FM baseline.
    \item Drug-proteins-disease, using GCE layer with FM baseline (FM-GCE).
    \item Drug-proteins-disease with side-effects as feature vectors of drug nodes, using GCE layer with FM baseline and side-information (FM-GCE-SI).
\end{enumerate}

\vspace{-1em}
\subsection{Analysis}

In our experiments, we use GCE in MF, FM and NFM. In all cases, the performance of the models improves when the GCE layers are used. This demonstrates that introducing GCE layers leads to statistically significant improvements with respect to the corresponding baseline models. We analyze the performance on long-tail users and items to see whether the GCE layer can preserve it's benefits on long tail users or items~\cite{zhang2020graph}. We perform the experiments in terms of NDCG@10 metric for FM baseline, and FM-GCE model on two datasets: ML-100K and User-Books interaction.

\subsubsection{Long-tail Items}

As shown in \autoref{table:analysis}(a), for both without and with GCE layer, performance on NDCG@10 drops a lot when subtracting the K popular items. This might be because popular items dominate the top of recommendation
lists and also due to the little amount of relevant neighbours that those items may have
. However, the proposed GCE layer achieves more significant improvements when recommending less popular items - the improvement increases the more popular items we subtract.
\subsubsection{Long-tail Users} Similarly, as shown in \autoref{table:analysis}(b), we sort users by the total of interactions they had and report the performance results when removing the k-popular users. We observe a trend that the relative improvement of the baseline methods when using the GCE layers is larger when removing popular items. This shows the potential of the GCE layer to improve the performance of baseline models in settings where less information is available, by propagating information over the N-partite graph. 
On the user side, we observe less of this effect and note that GCE maintains it relative performance with respect to the baselines models in most settings.

\subsection{Inductive bias effect}

Inductive bias are the assumptions encoded in a model to generalize its performance. In the case of a Graph Convolutional Embeddings the inductive bias is called {\bf relational} and it is based on the assumption that elements that are connected in the graph must have similar representations. This inductive bias in GCE is essentially encoded by the adjacency matrix of the model.

{\def\arraystretch{0.55}
\begin{table}[H]
\centering
{\begin{tabular}{L{1.2cm}C{1.4cm}C{1.15cm}C{1.15cm}C{2.1cm}}
    \toprule
    & 
    \textbf{ML-100K} & \textbf{Books} &\textbf{LastFM} & \textbf{Drugs-Disease} \\
    \midrule
    \gb\texttt{MF} & \gb 0.0302  & \gb 0.0189 & \gb 0.1594 & \gb 0.0320  \\
    \texttt{MF-GCE} & \textbf{0.0647}  & \textbf{0.0369} & \textbf{0.2150} & \textbf{0.1594}  \\
    \midrule
    \gb\texttt{FM} & \gb 0.0719  & \gb 0.0410 & \gb 0.2775 & \gb 0.1649  \\
    \texttt{FM-GCE} & \textbf{0.0872}  & \textbf{0.0548} & \textbf{0.3075} & \textbf{0.1905}  \\
    \midrule
    \gb\texttt{NCF} & \gb 0.0741 & \gb 0.0126 & \gb 0.2600 & \gb 0.1795  \\
    \texttt{NCF-GCE} & \textbf{0.0877}  & \textbf{0.0666} & \textbf{0.2701} & \textbf{0.1857}  \\
    \arrayrulecolor{black}\bottomrule
\end{tabular}}
\caption{First-epoch recommendation performance in terms of NDCG@10 metric for all baseline models in all datasets.}
\label{table:convergence}
\end{table}
}

\vspace{-1.5em}
\autoref{table:convergence} shows the model performance in ML-100k and User-Books Interaction datasets\footnote{The results in other datasets are similar and are omitted due to the space limitation.} for each of the baselines, after only one iteration of the training process with a random mini-batch of data. This experiment is designed to show the  effect of GCE's inductive bias with regard to performance. We observe that GCE layer allows the model to achieve significative better results than baseline models when evaluated after one training step, thus leading to consistently better performance. The GCE layer by definition encodes the graph topology (i.e. user-item-context interaction) information into the node embedding vectors, while conventional embeddings are initialized randomly. In the later case, gradient updates are the only source of information about interaction patterns as well as more complex information like the nodes' intrinsic features. Embeddings from the proposed GCE layer represent the interaction patterns explicitly, hence the training process is decoupled and the model can learn more complex signals. 

We also observe that the dataset structure needs to be consistent in order to apply GCE layer successfully. For instance, if the dataset structure concentrates the majority of the interactions on some specific users, or if the majority of items are clicked by just one user, that can lead to no improvement when applying the GCE layer. In those situations, some users seen as outliers can be removed or top-frequent items can be selected, respectively.


\subsection{Discussion}
\label{sec:discussion}
FM have been the reference models in the area of context-aware recommendation models. However, they do not model effectively high-order and nonlinear interaction signals. Several recent efforts focused on enriching FM with neural networks, such as NFM \cite{nfm} and NCF\cite{he2017neural} models, which are considered strong baselines. 

Additionally, \citet{rendle2020neural} showed that, by carefully optimizing the training process, expressive models such as NCF were not needed to outperform the classical MF. However, in our case, where context dimension are introduced, we do not observe this pattern. This can be due to the fact that MF is extended to tensor factorization in order to be adapted for context-aware recommendation. In fact, we observe that it is more efficient to compute pair-wise operations (like FM does) and one of the reasons might be that the number of parameters becomes too high when computing inner product along more than 2 dimensions. What can be observed in any of the datasets is that FM performance is much closer in terms of performance to NCF results, hence this might corroborate the idea of \cite{rendle2020neural} but for more than just user-item dimensions.

Additionally, it has to be noted that applying GCE layer on MF leads to more notable improvements, as it helps it to capture high-order interactions between the various dimensions of the context and user-item interactions as effectively as in other DL-based models. This further demonstrates that the proposed GCE layer increases even more the performance of downstream recommendation models that are not so strong acting as a potential equalizer of performance among recommendation models.

\section{Conclusion}
The main goal of this work was to ease the use of GCNs in the area of RS by building a GCN model that can be used as an embedding layer in practically any downstream recommendation model, processing generalized user-item-context interactions. 
In this paper, we achieved two contributions: 1) we generalized GCNs in $N$-partite interaction graphs to take context variables into account and 2) we integrated the generated embeddings in downstream recommendation models trained in an end-to-end fashion.

We presented an effective model for capturing high-order and non-linear interaction signals in context-aware recommender systems, which we call Graph Convolutional Embeddings\footnote{We release code, models and data in \url{https://drive.google.com/uc?id=1zaDoWMpyYGuvBUNdaz06XoKZcQO9Yjrb&export=download}}.
This architecture is based on the message passing principle of GCN, it is scalable and can be used with any deep learning recommendation model that naturally allows to incorporate context, such as FM, NCF, DFM, etc. Using our layer allows to easily incorporate side-information for each of the nodes. 

From the experimental results, we observed that using graph convolutions helps to capture patterns in the contextual data and thus leads to better model performance, especially on relatively sparse and large datasets. This is due to the fact that graph convolutions encode in each node embedding, not just the information of the node itself but also the information from correlated nodes; thus leveraging the graph topology and connectivity information contained by the constructed interaction graph.



\bibliographystyle{ACM-Reference-Format}
\bibliography{sample}

\end{document}